\documentstyle[sprocl]{article}

\input{epsf}
\newcommand{\binom}[2]{
\renewcommand{\arraystretch}{0.75}
\left(\begin{array}{@{}c@{}}#1\\#2\end{array}\right)
\renewcommand{\arraystretch}{1}
}

 \newcommand{\zr}[1]{\mbox{\hspace*{#1em}}}

 \newcommand{\ZZ}{\mbox{\sf Z\zr{-0.42}Z}}

\begin{document}

\title{COORDINATION SEQUENCES AND CRITICAL POINTS}

\author{\sc Michael Baake}

\address{Institut f\"ur Theoretische Physik, Universit\"at T\"ubingen, \\
Auf der Morgenstelle 14, D-72076 T\"ubingen, Germany}

\author{\sc Uwe Grimm, Przemys{\l}aw Repetowicz}

\address{Institut f\"ur Physik, Technische Universit\"{a}t, 
D-09107 Chemnitz, Germany}

\author{\sc Dieter Joseph}

\address{Laboratory of Atomic and Solid State Physics, \\ 
Cornell University, Ithaca, NY 14853-2501, USA }

\maketitle\abstracts{
Coordination sequences of periodic and quasiperiodic graphs
are analysed. These count the number of points that can be reached
from a given point of the graph by a number of steps along its bonds, 
thus generalising the familiar coordination number which is 
just the first member of this series. A possible application to the
theory of critical phenomena in lattice models is outlined.}

\section{Introduction}

The locations of critical points of lattice models (such as
Ising models, percolation problems or self-avoiding walks) depend
on the topological structure of the underlying graph, in particular
on the dimension and on the (mean) coordination number.$^{1-5}$
However, this can only be a first approximation, and
more detailed information on the graph is required for
a better understanding of the precise location of critical points.
For this purpose, quasiperiodic graphs prove helpful 
because they provide different cases with the same mean coordination 
number and also a check of the ``universality'' of this approach.

Below, we display first results on the combinatorial
issue to calculate the coordination sequences, and on general tendencies
visible. To this end, we summarise the results for root graphs$\,^{6,7}$ 
and present the averaged coordination numbers
for the rhombic Penrose and for the Ammann-Beenker tiling. 

\section{Coordination sequences}

A natural family of periodic graphs is given by the so-called root lattices
and their relatives, the root graphs,$^{7}$
encoded by the Dynkin
diagrams of $A_n$, $B_n$ ($n\geq 2$), $C_n$ ($n\geq 3$), $D_n$ ($n\geq 4$), 
$G_2$, $F_4$, $E_6$, $E_7$, and $E_8$.
In all these cases, the coordination sequences $s(k)$, 
where $s(k)$ is the number of $k$-th neighbours and $s(0):=1$,
can be calculated completely 
and encapsulated in an ordinary generating function of the form
\begin{equation}\label{genfun}
  S(x) \; = \; \sum_{k=0}^{\infty}\, s(k)\, x^k 
       \; = \; \frac{P(x)}{(1-x)^n}
\end{equation}
where $n$ is the dimension of the lattice (resp.\ the graph) and $P(x)$
is an integral polynomial of degree $n$. For the (hyper-)cubic 
lattice $\ZZ^n$, this polynomial is simply $P_n(x) = (1+x)^n$. 
For the other cases, the result is
\begin{equation}
\begin{array}{@{}r@{\; =\;}l@{\quad\;\;}r@{\; =\;}l@{}}
P_{A_n}^{}(x) & \displaystyle
\sum_{k=0}^{n}\, {\binom{n}{k}}^{\!\!\! 2} x^k\, , &
P_{B_n}^{}(x) & \displaystyle\sum_{k=0}^{n}\,\left[
\binom{2n+1}{2k} - 2k\binom{n}{k}\vphantom{\frac{2k(n-k)}{n-1}}
\right] x^k\, , \\[4ex]
P_{C_n}^{}(x) & \displaystyle\sum_{k=0}^{n}\binom{2n}{2k} x^k\, , &
P_{D_n}^{}(x) & \displaystyle\sum_{k=0}^{n}\,\left[
\binom{2n}{2k} -\frac{2k(n-k)}{n-1}\binom{n}{k}\right] x^k\, ,
\end{array}
\end{equation}
the expressions for the exceptional cases can be found in Refs.~6 and 7.
Many other cases can be calculated, in particular periodic graphs with
more complicated fundamental domains, where an averaging procedure over
the possible starting points is necessary. Details on this will be given
elsewhere.

\begin{table}[tb]
\vspace*{-2mm}
\caption{First coordination numbers of the Penrose tiling
[$\tau=(1+\protect\sqrt{5})/2$].}
\vspace{0.5ex}
\begin{center}
\begin{tabular}{|r|r@{$\;$}c@{$\;$}r|r@{.}l||r|r@{$\;$}c@{$\;$}r|r@{.}l|}
\hline
\multicolumn{1}{|c|}{\rule[-1ex]{0pt}{3.5ex}$k$\rule[-1ex]{0pt}{3.5ex}} & 
\multicolumn{3}{c|}{$s(k)$} & 
\multicolumn{2}{c||}{num.\ value} & 
\multicolumn{1}{c|}{$k$} &
\multicolumn{3}{c|}{$s(k)$} & 
\multicolumn{2}{c|}{num.\ value}\\
\hline
\rule[-0.5ex]{0ex}{3ex}
$1$ \rule[-0.5ex]{0ex}{3ex}
& $4$&& & $4$&$000\, 000$ &
   \rule[-0.5ex]{0ex}{3ex}
   $6$ \rule[-1ex]{0ex}{3ex} 
   & $980$&$-$&$588\tau$ & $28$&$596\, 015$\\
\rule[-0.5ex]{0ex}{2ex}
$2$ \rule[-0.5ex]{0ex}{2ex}
& $58$&$-$&$30\tau$ & $9$&$458\, 980$ &
   \rule[-0.5ex]{0ex}{2ex}
   $7$ \rule[-0.5ex]{0ex}{2ex}
   & $-1614$&$+$&$1018\tau$ & $33$&$158\, 601$ \\
\rule[-0.5ex]{0ex}{2ex}
$3$ \rule[-0.5ex]{0ex}{2ex}
& $-128$&$+$&$88\tau$ & $14$&$386\, 991$ &
   \rule[-0.5ex]{0ex}{2ex}
   $8$ \rule[-0.5ex]{0ex}{2ex} 
   & $2688$&$-$&$1638\tau$ & $37$&$660\, 326$ \\
\rule[-0.5ex]{0ex}{2ex}
$4$ \rule[-0.5ex]{0ex}{2ex}
& $288$&$-$&$166\tau$ & $19$&$406\, 358$ &
   \rule[-0.5ex]{0ex}{2ex}
   $9$ \rule[-0.5ex]{0ex}{2ex}
   & $-3840$&$+$&$2400\tau$ & $43$&$281\, 573$ \\
\rule[-1ex]{0ex}{2.5ex}
$5$ \rule[-1ex]{0ex}{2.5ex}
& $-374$&$+$&$246\tau$ & $\: 24$&$036\, 361$ &
   \rule[-1ex]{0ex}{2.5ex}
   $10$ \rule[-0.5ex]{0ex}{2.5ex}
   & $4246$&$-$&$2594\tau$ & $\: 48$&$819\, 833$ \\
\hline
\end{tabular}
\end{center}
\vspace*{-2.5mm}
\end{table}

In view of the applications, one is interested in a variety of graphs
that share the same mean coordination number, but have different (averaged)
higher coordination numbers. Such examples are provided by planar rhombic
tilings with $N$-fold rotational symmetry. As they are solely built 
from rhombi, the average coordination number is always exactly four, but 
the second (and higher)
numbers may vary. To expand on this, we present the coordination numbers
for two examples, the rhombic Penrose and the Ammann-Beenker tiling (also
known as the standard octagonal tiling). 
The values for the Penrose tiling given in Table~1 were obtained by an exact
window algorithm, whereas the results for the Ammann-Beenker case in Table~2 
are conjectures based on the averaged coordination numbers of a periodic 
approximant with 275807 vertices.

\begin{table}[tb]
\vspace*{-2mm}
\caption{First coordination numbers of the octagonal tiling 
($\lambda=1+\protect\sqrt{2}$).}
{\small
\vspace{0.5ex}
\begin{center}
\begin{tabular}{|r|r@{$\;$}c@{$\;$}r|r@{.}l||r|r@{$\;$}c@{$\;$}r|r@{.}l|}
\hline
\multicolumn{1}{|c|}{\rule[-1ex]{0pt}{3.5ex}$k$\rule[-1ex]{0pt}{3.5ex}} & 
\multicolumn{3}{c|}{$s(k)$} & 
\multicolumn{2}{c||}{num.\ value} & 
\multicolumn{1}{c|}{$k$} &
\multicolumn{3}{c|}{$s(k)$} & 
\multicolumn{2}{c|}{num.\ value}\\
\hline
\rule[-0.5ex]{0ex}{3ex}
$1$ \rule[-0.5ex]{0ex}{3ex}
& $4$&& & $4$&$000\, 000$ &
   \rule[-0.5ex]{0ex}{3ex}
   $11$ \rule[-1ex]{0ex}{3ex} 
   & $360$&$-$&$128\lambda$ & $50$&$980\, 664$\\
\rule[-0.5ex]{0ex}{2ex}
$2$ \rule[-0.5ex]{0ex}{2ex}
& $48$&$-$&$16\lambda$ & $9$&$372\, 583$ &
   \rule[-0.5ex]{0ex}{2ex}
   $12$ \rule[-0.5ex]{0ex}{2ex}
   & $828$&$-$&$320\lambda$ & $55$&$451\, 660$ \\
\rule[-0.5ex]{0ex}{2ex}
$3$ \rule[-0.5ex]{0ex}{2ex}
& $-24$&$+$&$16\lambda$ & $14$&$627\, 417$ &
   \rule[-0.5ex]{0ex}{2ex}
   $13$ \rule[-0.5ex]{0ex}{2ex} 
   & $-508$&$+$&$236\lambda$ & $61$&$754\, 401$ \\
\rule[-0.5ex]{0ex}{2ex}
$4$ \rule[-0.5ex]{0ex}{2ex}
& $28$&$-$&$4\lambda$ & $18$&$343\, 146$ &
   \rule[-0.5ex]{0ex}{2ex}
   $14$ \rule[-0.5ex]{0ex}{2ex}
   & $-996$&$+$&$440\lambda$ & $66$&$253\, 967$ \\
\rule[-0.5ex]{0ex}{2ex}
$5$ \rule[-0.5ex]{0ex}{2ex}
& $52$&$-$&$12\lambda$ & $23$&$029\, 437$ &
   \rule[-0.5ex]{0ex}{2ex}
   $15$ \rule[-0.5ex]{0ex}{2ex}
   & $2580$&$-$&$1040\lambda$ & $69$&$217\, 895$ \\
\rule[-0.5ex]{0ex}{2ex}
$6$ \rule[-0.5ex]{0ex}{2ex}
& $48$&$-$&$8\lambda$ & $28$&$686\, 282$ &
   \rule[-0.5ex]{0ex}{2ex}
   $16$ \rule[-0.5ex]{0ex}{2ex}
   & $1620$&$-$&$640\lambda$ & $74$&$903\, 320$ \\
\rule[-0.5ex]{0ex}{2ex}
$7$ \rule[-0.5ex]{0ex}{2ex}
& $-324$&$+$&$148\lambda$ & $33$&$303\, 607$ &
   \rule[-0.5ex]{0ex}{2ex}
   $17$ \rule[-0.5ex]{0ex}{2ex}
   & $-5288$&$+$&$2224\lambda$ & $81$&$210\, 963$ \\
\rule[-0.5ex]{0ex}{2ex}
$8$ \rule[-0.5ex]{0ex}{2ex}
& $732$&$-$&$288\lambda$ & $36$&$706\, 494$ &
   \rule[-0.5ex]{0ex}{2ex}
   $18$ \rule[-0.5ex]{0ex}{2ex}
   & $2372$&$-$&$948\lambda$ & $83$&$325\, 543$ \\
\rule[-0.5ex]{0ex}{2ex}
$9$ \rule[-0.5ex]{0ex}{2ex}
& $380$&$-$&$140\lambda$ & $42$&$010\, 101$ &
   \rule[-0.5ex]{0ex}{2ex}
   $19$ \rule[-0.5ex]{0ex}{2ex}
   & $1324$&$-$&$512\lambda$ & $87$&$922\, 656$ \\
\rule[-1ex]{0ex}{2.5ex}
$10$ \rule[-1ex]{0ex}{2.5ex}
& $-1140$&$+$&$492\lambda$ & $\: 47$&$793\, 073$ &
   \rule[-1ex]{0ex}{2.5ex}
   $20$ \rule[-1ex]{0ex}{2.5ex}
   & $1196$&$-$&$456\lambda$ & $\: 95$&$118\, 616$ \\
\hline
\end{tabular}
\end{center}
}
\vspace*{-2.5mm}
\end{table}

\section{Critical points}

So far, only rudimentary results exist. Critical points of
periodic systems have been investigated in some detail, but the results
are not fully convincing, and they are still rather incomplete. Furthermore,
most results concern hypercubic lattices and hence graphs with different
dimensions and coordination numbers, which obscures the dependence on higher
order coordination numbers.

{}For the problem of self-avoiding walks, there exist results also on a
variety of quasiperiodic tilings,$^{2}$ and their analysis supports
our claim that higher order coordination numbers move the location.
In particular, it is clearly seen that the critical point increases with
growing second coordination number, while the critical exponents seem
to be universal, as expected. However, 
in order to arrive at a conclusive statement,
more, and in particular more precise, data on graphs with coinciding
dimension and mean coordination number are needed.

\section{Outlook}

It is an open question whether one can find explicit expressions for
the generating functions $S(x)$ for quasiperiodic graphs. Clearly, these
will not have the form of Eq.~(\ref{genfun}) --- an analysis of the
generating functions of periodic approximants shows that the polynomial 
degrees of numerator and denominator grow with the size of the approximant.
Still, the coordination sequences show quite a regular behaviour, 
compare Fig.~1, also when compared to slightly more complicated periodic
graphs as, for instance, the Kagom\'{e} and the diced lattice. The latter
show some similarity to the (dual) Penrose tiling with regard
to high-temperature expansions.$^{8}$
{Their mean coordination numbers $s(k)$ grow linearly with $k$, 
but with two different slopes for even and for odd values of $k$.} 

\begin{figure}[bt]
\centerline{\epsfxsize=0.765\textwidth\epsfbox{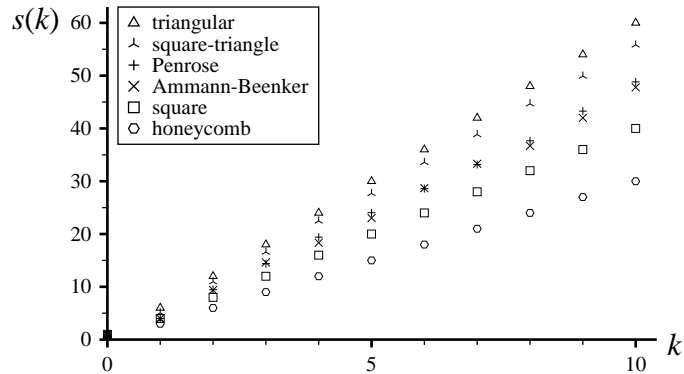}}\vspace*{-3ex}
\caption{Coordination sequences of some periodic and quasiperiodic 
graphs.\label{fig1}}
\vspace*{-2.5mm}
\end{figure}

Ideally, one would hope for a kind of integral transform from the generating
function of the combinatorial problem to the partition sum of the 
corresponding model of statistical mechanics. Such a transformation, 
however, cannot exist in
general since these models are usually not solvable, so their
partition sums have an analytic behaviour incompatible with that of our 
generating functions.

Therefore, one can only expect an empirical formula for the correction to the
location of critical points due to higher order coordination numbers. To get
a better understanding of this, we intend to analyse the critical structure
of various models and on a variety of graphs in the near future.

\end{document}